\documentclass[prl,aps,a4paper,twocolumn,footinbib]{revtex4}

\newif\ifusesec
\usesecfalse

\usepackage{graphicx,psfrag}
\usepackage{mathrsfs}
\usepackage{amsmath,amsfonts,amssymb}
\usepackage{multirow}
\usepackage{comment}
\usepackage{color}
\usepackage{hyperref}

\def\p{{\partial}}

\newcommand{\be}{\begin{equation}}
\newcommand{\ee}{\end{equation}}

\definecolor{cyan}{rgb}{0,0.9,0.9}
\definecolor{orange}{rgb}{0.9,0.5,0}
\definecolor{magenta}{rgb}{1,0,1}
\definecolor{purple}{rgb}{0.8,0.4,0.8}
\definecolor{gray}{rgb}{0.8242,0.8242,0.8242}

\begin{document}

\title{Quasiuniversal properties of neutron star mergers}
 
\author{Sebastiano \surname{Bernuzzi}$^1$}
\author{Alessandro \surname{Nagar}$^2$}
\author{Simone \surname{Balmelli}$^3$}
\author{Tim \surname{Dietrich}$^1$}
\author{Maximiliano \surname{Ujevic}$^{4}$}
\affiliation{$^1$Theoretical Physics Institute, University of Jena, 07743
  Jena, Germany}   
  \affiliation{$^2$Institut des Hautes Etudes Scientifiques, 91440
  Bures-sur-Yvette, France} 
\affiliation{$^3$Physik-Institut, Universit\"at
  Z\"urich, 8057 Z\"urich, Switzerland} 
\affiliation{$^4$Centro de Ci\^encias Naturais e Humanas, Universidade Federal do ABC,
09210-170, Santo Andr\'e, S\~ao Paulo, Brazil}

\date{\today}

\begin{abstract}
  Binary neutron star mergers are studied using nonlinear 3+1
  numerical relativity simulations and the analytical
  effective-one-body (EOB) model.   
  The EOB model predicts quasiuniversal relations between the
  mass-rescaled gravitational wave frequency and the binding energy at
  the moment of merger, and certain dimensionless binary tidal coupling
  constants depending on the stars Love numbers, compactnesses and the binary 
  mass ratio.
  These relations are quasiuniversal in the sense that, for a given
  value of the 
  tidal coupling constant, they depend significantly neither on the
  equation of state nor on the mass ratio, though they do depend on
  stars spins.   
  The spin dependence is approximately linear for small spins aligned 
  with the orbital angular momentum. 
  The quasiuniversality is a property of the conservative dynamics; 
  nontrivial relations emerge as the binary interaction becomes
  tidally dominated. 
  This analytical prediction is qualitatively consistent with new,
  multi-orbit numerical relativity results for the relevant case of
  equal-mass irrotational binaries. 
  Universal relations are thus expected to characterize neutron star
  mergers dynamics.   
  In the context of gravitational wave astronomy, these universal relations may
  be used to constrain the neutron star equation of state using
  waveforms that model the merger accurately. 
\end{abstract}


\maketitle

\ifusesec
\section{Introduction}
\else
\paragraph*{Introduction.--}
\fi
\label{sec:intro}
Binary neutron star (BNS) inspirals are among the most promising sources for the 
advanced configurations of the ground based gravitational wave (GW)
detector network~\cite{Abadie:2010cf}. 
Advanced configurations of LIGO and Virgo detectors are expected to
listen to $\sim0.4-400$~yr$^{-1}$ events starting from
2016-19~\cite{Aasi:2013wya}. Direct GW observations will then probe
such systems in the near future. In particular, because the 
late--inspiral--merger phase depends crucially on the stars internal
structure, the measurement of the tidal polarizablity parameters from
GWs will put the strongest constraints on the unknown nuclear equation of 
state~(EOS)~\cite{Read:2009yp,Damour:2012yf,Read:2013zra,DelPozzo:2013ala}.

An accurate modeling of neutron star mergers requires numerical relativity (NR).
In recent years simulations have become fairly robust, but exploring 
the physical parameter space remains a challenge out of reach. Furthermore, 
the interpretation of simulation data can be nontrivial: meaningful quantities must be 
gauge invariant and possibly have well-defined post-Newtonian (PN) limits.
The GW phasing analysis for multi-orbits ($\sim 10$) simulations was performed
by some groups, e.g.~\cite{Baiotti:2010xh,Bernuzzi:2011aq,Hotokezaka:2013mm}.
The BNS dynamics, expressed via the gauge-invariant relation between
binding energy and angular momentum~\cite{Damour:2011fu,Taracchini:2013rva}, 
was recently analyzed in both the nonspinning and spinning 
case~\cite{Bernuzzi:2012ci,Bernuzzi:2013rza}. For both observables a
solid analytical framework, although approximate, is essential for
extracting information from the simulations. 

Despite these detailed studies, simple, and
fundamental questions about the merger physics still lack of
quantitative answers. 
For instance, a test-mass in the Schwarzschild metric of mass $M$ has
a last stable orbit (LSO) at $R_{\rm LSO}=6M$, (we use units with
$G=c=1$) with dimensionless (or mass-reduced) orbital frequency $M\Omega_{\rm LSO}^{\rm
  Schw}=6^{-3/2}\approx 0.06804$. 
The associated GW frequency $2 M\Omega_{\rm LSO}^{\rm Schw}\approx
0.13608$ is commonly used to mark the end of the quasiadiabatic BNS
inspiral, setting $M$ equal to the total mass of the binary.  
Similarly,  the specific LSO binding energy $E_{b\, \rm LSO}^{\rm Schw}=(8/9)^{1/2}-1\approx-0.0572$ is used  
to estimate the total amount of GW energy emitted during the coalescence
process.  
These numbers appear ubiquitously in BNS-related studies,
e.g.,~\cite{DelPozzo:2013ala}, but are, in principle, no more 
then an order of magnitude estimates as they neglect both finite
mass ratio and finite size effects. 
Some questions arise:
How to model/include these effects? 
How does the merger frequency and binding energy depend on the main
parameters of the binary (EOS, mass ratio and individual spins)?  
How accurate are the Schwarzschild LSO estimates?
In this work we use new multi-orbit NR data and the analytical
effective-one-body (EOB) approach 
problem to put forward some answers. 
We find that the GW frequency and binding energy at the moment of
merger are characterized only by certain dimensionless tidal coupling 
constants (a fact also empirically observed in~\cite{Read:2013zra} for the
frequency) and the stars spins as a consequence of a fundamental
property of the underlying conservative dynamics.

\ifusesec
\section{EOB and the LSO}
\else
\paragraph*{EOB and the LSO.--}
\fi
\label{sec:eob}
The EOB
formalism~\cite{Buonanno:1998gg,Buonanno:2000ef,Damour:2000we,Damour:2001tu}
maps the relativistic 2-body problem, with masses $M_{A}$ 
and $M_{B}$, into the motion of an effective particle of mass
$\mu=M_{A}M_{B}/M$, with $M=M_{A}+M_{B}$, moving into an effective
metric. It employs standard PN results (e.g.,~\cite{Blanchet:2006zz}) in a
{\it resummed} form, and it is robust and predictive also in the
strong-field and fast-motion regime. The EOB model can be completed
with NR information; complete (inspiral-merger-ringdown) binary black hole waveforms for GW
astronomy can be produced for general mass-ratio and spin
configurations~\cite{Damour:2012ky,Taracchini:2013rva}.  
Tidal effects can also be included in the model~\cite{Damour:2009wj,Damour:2012yf}.
The EOB model consists of three building blocks: 
(i)~a Hamiltonian $H_{\rm EOB}$; 
(ii)~a factorized gravitational waveform; 
and (iii)~a radiation reaction force ${\cal F}_{\varphi}$.
The EOB Hamiltonian is $H_{\rm EOB} = M\sqrt{1+2\nu(\hat{H}_{\rm eff}-1)}$
where, in the nonspinning case, 
$ \hat{H}_{\rm eff}(u,p_{r_*},p_\varphi)\equiv H_{\rm eff}/\mu = 
\sqrt{A(u;\nu)\,(1  + p^2_\varphi u^2 + 2\nu(4-3\nu)u^2 p_{r*}^4) + p_{r*}^2} $, 
with $\nu\equiv \mu/M$, $u\equiv 1/r\equiv GM/R c^{2}$, $p_\varphi \equiv P_{\varphi}/(M\mu)$
is the dimensionless orbital angular momentum and $p_{r_*}\equiv
\sqrt{A/B}p_r=P_r/\mu$ is a dimensionless radial momentum, $A(u;\,\nu)$ and 
$B(u;\,\nu)$ are the EOB potentials.  
The conservative dynamics (${\cal F}_{\varphi}=0$) along circular
orbits ($p_{r_*}=0$) is determined only by $A(u;\,\nu)$.
Finite-size effects are formally 5PN. They are included in $A(u;\,\nu)$ by
adding a tidal term $A^T(u;\,\nu)$ to the point-mass $A^0(u;\,\nu)$ 
contribution, i.e.~$A(u)\equiv A^0(u;\nu)+A^T(u;\nu)$~\cite{Damour:2009wj}. 
The $A^{0}(u)$ function is analytically known at 4PN accuracy and formally reads
\hbox{$A^{0}_{\rm 4PN}(u;\nu)=1-2u+ \nu \hat{a}_{\rm 4PN}(u;\, \nu)$},
where $\hat{a}_{\rm 4PN}(u;\, \nu) \equiv a_{3} u^{3} + a_{4}u^{4} +
(a_{5}^{c}(\nu)+a_{5}^{\ln}\ln u) u^{5}$~\cite{Bini:2013zaa}. 
We use here {\it only} the 4PN-accurate analytical information 
and we do not add any ``flexibility parameter'' calibrated to NR data.
The Taylor-expanded function $A^{0}_{\rm 4PN}$ is resummed using
a $(1,4)$ Pad\'e approximant, i.e.~ $A^0(u;\nu)\equiv P^1_4[A^{0}_{\rm
    4PN}(u;\nu)]$,  
with the logarithmic term treated as a constant in the Pad\'e. 
The tidal part of the interaction potential is known at
next-to-next-to-leading order (NNLO, fractional 2PN) and reads
$A^T(u)=-\sum_{\ell=2}^4 \kappa^T_\ell
u^{2\ell+2}(1+\bar{\alpha}^{(\ell)}_1 
u+\bar{\alpha}^{(\ell)}_2 u^2)$, with only $\bar{\alpha}^{(2),(3)}_{1,2}$
known analytically~\cite{Bini:2012gu}. For $\ell\geq 2$, the dimensionless 
tidal coupling constants are~\cite{Damour:2009wj} 
\be
\label{kappaT}
\kappa^T_\ell \equiv
2\left[\dfrac{1}{q}\left(\dfrac{X_{A}}{C_{A}}\right)^{2\ell+1}k^{A}_{\ell}+q\left(\dfrac{X_{B}}{C_{B}}\right)^{2\ell+1}k^{B}_{\ell}\right], 
\ee
where $q=M_{A}/M_{B}\geq1$, $X_{A}\equiv M_{A}/M=q/(1+q)$, $X_{B}\equiv
M_{B}/M=1/(1+q)$, $k_{\ell}^{A,B}$ and $C_{A,B}$ are the dimensionless
Love numbers and compactness of star $A$ and $B$.    
All the information about the EOS is encoded in the 
$\kappa_{\ell}^{T}$'s. 
For typical compactnesses $C\sim0.12-0.2$,
$\kappa^T_2\sim\mathcal{O}(10^2)$ and
$\kappa^T_{3,4}\sim\mathcal{O}(10^3)$.  

Stable circular orbits correspond to minima in $u$ of $\hat{H}_{\rm eff}$ for
a given value of $p_{\varphi}$. 
For any $u$, the condition $\hat{H}_{\rm eff}(u)'=0$
yields  $j^{2}(u)=-A'(u)/(u^{2}A(u))'$ for the angular momentum 
along circular orbits $j\equiv p_{\varphi}$ ($'\equiv\partial_u$). 
The orbital frequency reads
$ M\Omega(u;\nu) = \mu^{-1}\p_{j} H_{\rm EOB} 
 = j(u) A(u;\nu) u^2/(H_{\rm EOB} \hat{H}_{\rm eff})$.
The end of the adiabatic (circular) dynamics is marked by the LSO,
i.e., the inflection point of $\hat{H}_{\rm eff}$, that yields
$(u_{\rm LSO},j_{\rm LSO})$ and in turn the LSO orbital frequency
$M\Omega_{\rm LSO}(\nu)$. 
The Schwarzschild LSO frequency is recovered by construction
$M\Omega_{\rm LSO}(\nu=0)=M\Omega^{\rm Schw}_{\rm LSO}$. The
$\nu$-dependent, nontidal, corrections to $A$ 
are globally repulsive, 
i.e.,~$M\Omega_{\rm LSO}(\nu)>M\Omega^{\rm Schw}_{\rm LSO}$~\cite{Buonanno:1998gg}. 
The tidal contribution $A^{T}$ is, instead, always attractive, and
moves $M\Omega_{\rm LSO}$ to  lower frequencies. The LSO frequency
results then as a balance between repulsive and attractive
effects. 

Spin effects are included following Ref.~\cite{Damour:2001tu}, which
is robust enough for BNS realistic spin values (dimensionless
magnitude $\chi_{A,B}\lesssim 0.1$). The spin-orbit interaction is taken  
at NNLO~\cite{Nagar:2011fx}, the spin-spin at leading-order~\cite{Balmelli:2013PhRvD..87l4036B}. 
The spin gauge freedom is fixed according to~\cite{Damour:2008qf,Nagar:2011fx}.
To have circular orbits, we only consider spins parallel (or antiparallel) to the orbital angular 
momentum. The LSO computation is analogous to the nonspinning
case. $M\Omega_{\rm LSO}$ is larger (smaller) than the nonspinning case for
parallel (antiparallel) spins, i.e.,~the system is less (more)
bound~\cite{Damour:2001tu,Bernuzzi:2013rza}. 

The complete nonadiabatic EOB model (${\cal F}_{\varphi}\neq 0$)
allows one to go beyond the adiabatic-circular-LSO analysis and to
examine the model quantitatively with NR data.
For the radiation reaction $\cal{F}_\varphi$ we use the tidal
extension of the point-mass prescriptions of~\cite{Damour:2012yf}, and
also include a radial component
~\footnote{Contrarily to~\cite{Baiotti:2010xh,Bernuzzi:2012ci},
  NR-determined next-to-quasicircular corrections to the waveform
  and to $\cal{F}_\varphi$ are not included.}. 
The point-mass dynamics is taken at 4PN in both 
the $A(u;\nu)$ and $\bar{D}^0(u;\nu)\equiv [A(u;\nu) B(u;\nu)]^{-1}$ 
functions, using in the latter linear-in-$\nu$ 4PN coefficient obtained
numerically~\cite{Barack:2010ny,Barausse:2011dq}.  
Note that the formal regime of validity of the model may break when the
dynamics is evaluated for $u\gtrsim u_{\rm LSO}$ since the two stars
may be already in contact at those radial separations~\cite{Damour:2009wj}.

\begin{figure*}[t]
  \centering
  \includegraphics[width=.48\textwidth]{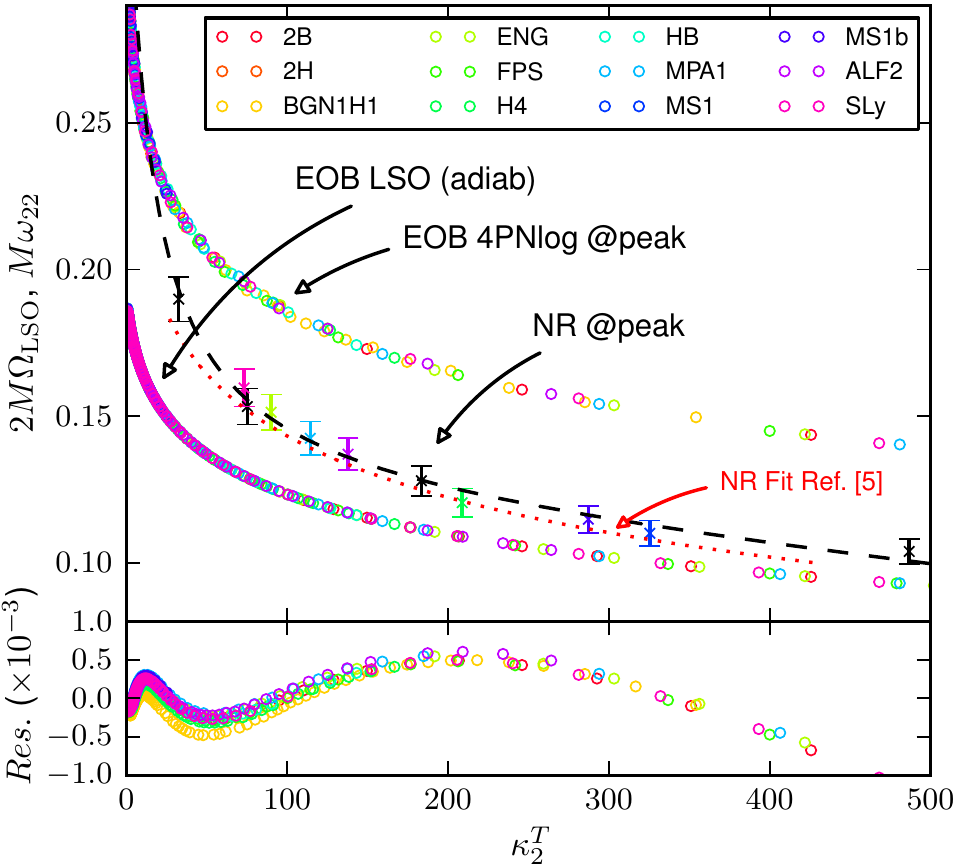} 
  \includegraphics[width=.48\textwidth]{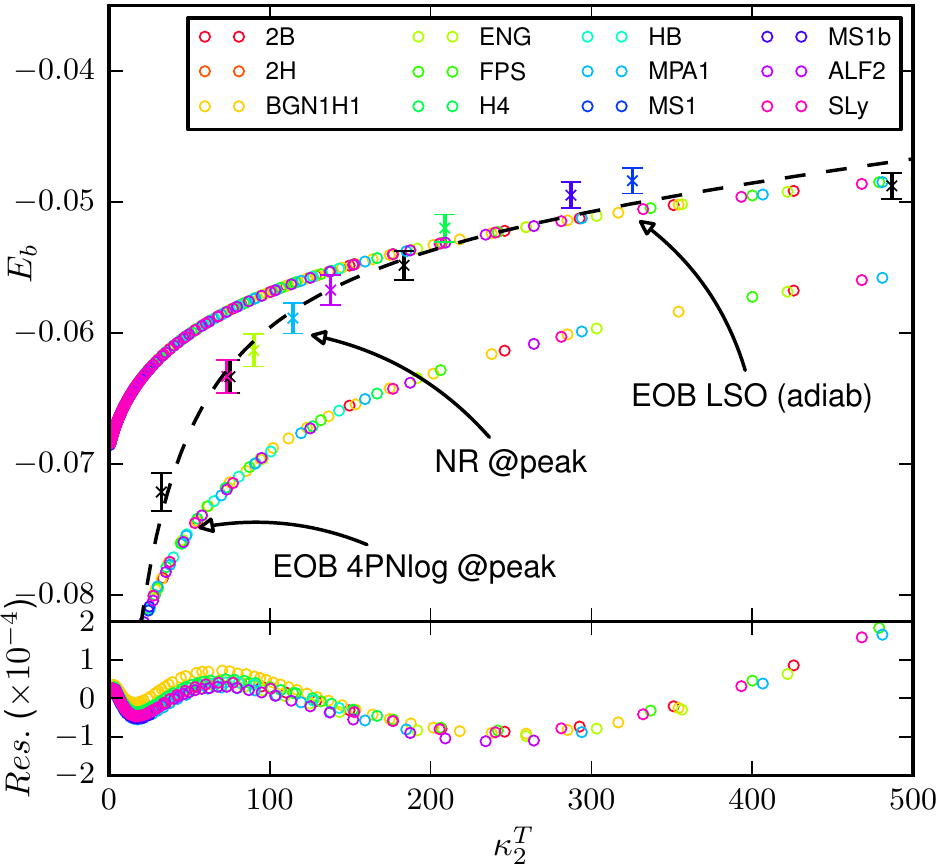}
    \caption{\label{fig:MoE} GW frequency (left) and
      binding energy (right) versus the coupling constant $\kappa^T_2$
      for equal-masses, irrotational mergers. 
      {\it Main panels:} %
      Circles refer to EOB quantities computed at either the
      adiabatic LSO ($2M\Omega_{\rm  LSO}$, $E_{b\, \rm LSO}$) 
      or the moment of merger ($M\omega_{22\, mrg}^{\rm EOB}$, $E_{b\
        mrg}^{\rm EOB}$). Different colors refer to different EOS.
      crosses (with error bars) refer to NR quantities 
      at the moment of merger,
      ($M\omega_{22\, mrg}^{\rm NR}$, $E_{b\, mrg}^{\rm NR}$). Among these, 
      the black crosses refer to polytropic EOS.
      The dashed black lines are the fits given in the text.
      The dotted red line in the left panel is the phenomenological
      fit of~\cite{Read:2013zra}. 
      {\it Bottom panels:} %
      Differences in $2M\Omega_{\rm
        LSO}$ and $E_{b\, \rm LSO}$ with respect the fits of the LSO
      data. An analogue result holds for the nonadiabatic EOB
      quantities at the moment of merger.
      The EOS dependence is negligible: all quantities (EOB LSO,
      4PNlog EOB, and NR) show $\kappa^T_\ell$-universality.}
\end{figure*}

\begin{figure}[t]
  \centering
    \includegraphics[width=.48\textwidth]{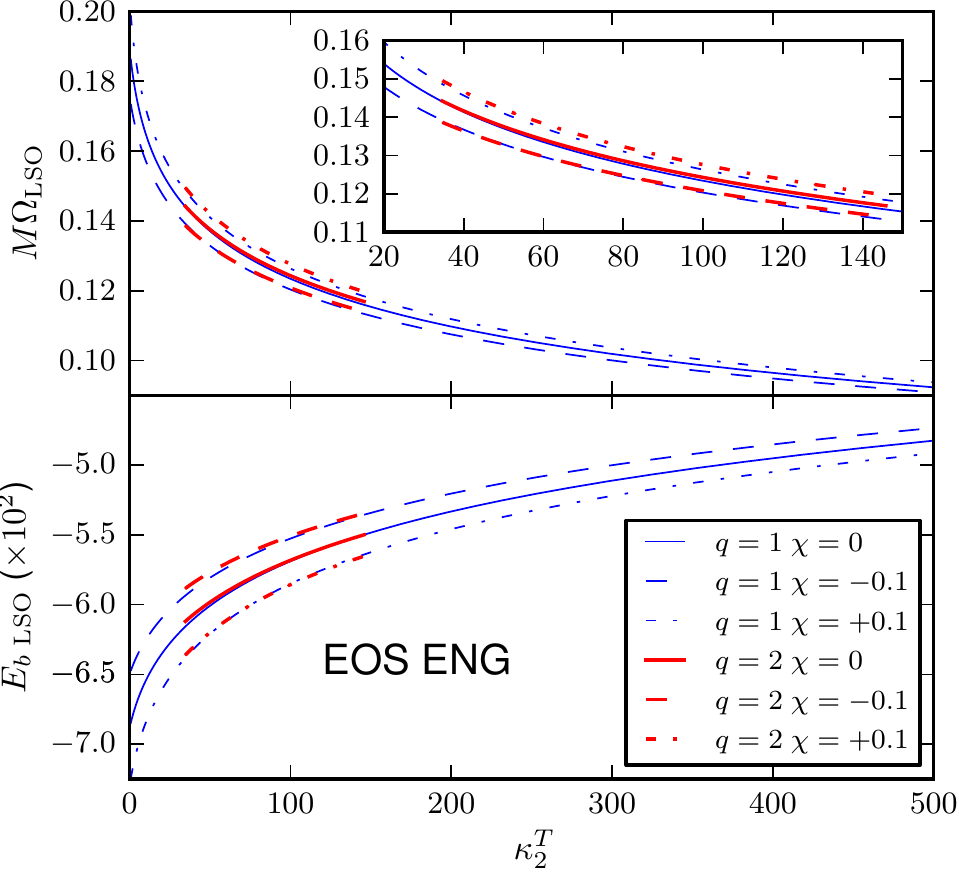}
    \caption{\label{fig:LSO_spin_and_q} 
      GW frequency (top) and
      binding energy (bottom) versus the coupling constant $\kappa^T_2$
      at EOB LSO: varying mass ratio and spin magnitude. Only the ENG EOS is
      plotted for simplicity.  
      The effect of mass ratio is almost negligible. The effect of
      spin is dominated by spin-orbit coupling.}  
\end{figure}

\ifusesec
\section{$\kappa^{T}_{\ell}$-universal relations}
\else
\paragraph*{$\kappa^{T}_{\ell}$-universal relations.--}
\fi
\label{sec:univ}
We studied the dependence of $2M\Omega_{\rm LSO}$ and the binding energy
per reduced mass at LSO, $E_{b\, \rm LSO}=(H_{\rm EOB}-M)/\mu$, when varying
EOS, compactness, mass ratio and spin. For each EOS in a sample of 12
realistic ones, we vary the mass of each star 
between $1.3M_{\odot}$ and the maximum mass allowed, $M_{\rm max}\gtrsim2M_\odot$.
We found that both $2M\Omega_{\rm LSO}$ and $E_{b\, \rm LSO}$ 
are essentially independent of the choice of EOS when expressed versus any
of the tidal coupling constant $\kappa^T_{\ell}$. 
For example, Fig.~\ref{fig:MoE} displays 
$2M\Omega_{\rm LSO}$ and $E_{b\, \rm LSO}$  versus the dominant 
coupling constant $\kappa_{2}^{T}$ for $q=1$ and no spins. 
From the residuals (bottom panels) 
one sees that deviations from universality are below the $0.2\%$. 
The same quasiuniversal behavior is found also for unequal-mass, spinning BNS. 
Varying $1\leq q\leq 2$ does not lead to curves significantly different 
from those in Fig.~\ref{fig:MoE}, the only difference
being a narrower interval of variability of $\kappa_{2}^{T}$. By contrast, the 
spin-orbit coupling, significantly changes the EOB LSO frequency and binding 
energy already at spin magnitudes $\chi\sim0.01-0.1$. 
An example is given by Fig.~\ref{fig:LSO_spin_and_q}, restricted to
EOS ENG for clarity. 
The dimensionless spin value is chosen to be  $\chi=\pm 0.1$. 
The difference between $q=2$ 
and $q=1$ curves is $\lesssim0.5\%$.  
The spin dependence is linear for spins $\chi\lesssim0.1$, as expected
for the spin-orbit interaction. 
Note that the functional dependence $2M\Omega_{\rm
  LSO}(\kappa^T_2)$ (and similarly $E_{b\, \rm LSO}(\kappa^T_2)$), is
algebraically complicated already for the simplest choice of the  
$A(u)$ function and cannot be made explicit.
Both quantities can be robustly fitted to a low-order rational
polynomial of the form $f(\kappa)=f(0)(1+n_1 \kappa + n_2 \kappa^2)/(1+d_1
\kappa + d_2 \kappa^2)$, where $f(0)$ is the point-mass LSO value
$(2M\Omega_{\rm  LSO}(0),E_{b\, \rm LSO}(0))\approx(0.1892,-0.0688)$.

As merger is approached, the dynamics enters a tidally dominated
regime: the values of $2M\Omega_{\rm LSO}$ and $E_{b\, \rm LSO}$ are
strongly influenced by tidal effects. 
Close to the LSO the tidal potential $A^T(u;\,\nu)$ may become
comparable or larger than $\nu\hat{a}(u;\,\nu)\equiv
A^{0}(u;\nu)-(1-2u)$, that determines point-mass ($\nu$-dependent)
effects. 
One can see this comparing the various 
contributions to the ``radial force''  
$dA/dr = -u^{2}\left[-2 + \nu\hat{a}'(u;\, \nu) +A'_{T}(u;\,\nu)\right]$. 
For example, at LSO (EOS SLy, $\nu=1/4$) one has:
for $C=0.14$ and $\kappa_{2}^{T}=274.51$, 
$u_{\rm LSO}\approx 0.1366$, 
which yield $\nu\hat{a}'(u)\approx0.0703$ and
$A'_{T}(u;\,\nu)\approx-0.1168$; 
for $C=0.18$ and $\kappa_{2}^{T}=58.52$, 
$u_{\rm LSO}\approx 0.1645$, 
which yield $\nu\hat{a}'(u)\approx0.1127$ and
$A'_{T}(u;\,\nu)\approx -0.0669$. 
Concerning the LSO frequency, one gets 
$M\Omega_{\rm LSO}=0.0517$ for $C=0.14$ and $M\Omega_{\rm
  LSO}=0.06674$ for $C=0.18$~\footnote{ 
  We stress that the result is qualitatively and quantitatively robust
  when changing the PN order of $A^0$ from 3PN to the 5PN  (the latter
  employs NR-tuned flexibility parameters~\cite{Damour:2012ky}). 
  We observe monotonic behavior with the various PN order. The fractional 
  difference between 5PN and 4PN is between 0.3 to 1\% for $\kappa_2^T\gtrsim50$, 
  and up to 3\% for $\kappa_2^T\in[10,50]$.}. 
The values of $M\Omega_{\rm LSO}$ and $E_{b\, \rm LSO}$ are rather
close to the Schwarzschild ones, being the latter determined by
$A(u;\,\nu=0)=1-2u$. 
The behavior is {\it not} a property of the LSO, but it is expected
to hold also for $u>u_{\rm LSO}$, since $A^T(u)\propto u^{6}$; i.e.~it holds
during the whole merger process.
By contrast, the universal curves extracted at separations
larger then the LSO progressively flatten (the
$\kappa^T_\ell$-dependency weakens) and approach the degenerate
point-mass case as the tidal interaction becomes negligible.

The complete nonadiabatic EOB dynamics can be continued
also after the LSO crossing and the orbital frequency
$M\Omega(t)$ develops a local maximum~\cite{Bernuzzi:2012ci}, likewise 
the point-mass case. The analytical time-domain $\ell=m=2$ EOB waveform 
is characterized by a peak in the modulus and a peak in the frequency
$M\omega^{\rm EOB}_{22}$,
reproducing the well-known qualitative structure of the NR 
waveforms, e.g.~\cite{Thierfelder:2011yi,Read:2013zra}. In this sense, 
the complete tidal EOB waveform qualitatively implements ``the merger'', 
already at the analytical level, i.e.~without NR-tuning. 
We define the {\it moment of merger} (in both EOB and NR)
as the peak of the amplitude of the $\ell=m=2$ mode of the GW. This is  
an idealization since the actual merger process takes place during 
the last few orbits of the coalescence. 
As shown in Fig.~\ref{fig:MoE}, the EOB wave frequency $M\omega^{\rm
  EOB}_{22\, mrg}$ and the binding
energy $E_{b\, mrg}^{\rm EOB}$  at the moment of merger are also
characterized by a $\kappa^T_\ell$-universality.

\ifusesec
\section{Comparison with NR}
\else
\paragraph*{Comparison with NR.--}
\fi
\label{sec:NR}
The adiabatic tidal EOB analysis captures the relevant qualitative 
features of the merger dynamics. 
Specifically, the quasiuniversal properties of $M\Omega$ and 
$E_b$ close to the EOB LSO hold also for the actual NR merger
frequency and binding energy.  
We stress that we do {\it not} advocate a formal link between the EOB
LSO and NR quantities, but  
rather give a suggestive argument for the existence of these universal
structures. 

We performed new NR simulations of coalescing BNS, employing the BAM
code and the method described in ~\cite{Brugmann:2008zz,Thierfelder:2011yi},
though: (i)~we use the Z4c formulation of Einstein's equations~\cite{Bernuzzi:2009ex}; 
(ii)~GWs are extracted from an extended wavezone~\cite{Hilditch:2012fp}.
The binaries are equal-mass, irrotational configurations with
different EOSs. A $\Gamma=2$ polytropic EOS model is employed to
simulate different compactnesses
$C_A=C_B=(0.12,\,0.14,\,0.16,\,0.18)$; EOS MS1, MS1b, H4, ALF2, MPA1,
ENG, SLy are employed for simulations with fixed isolation mass $M=2\times1.35M_\odot$. 
The evolutions covers about ten orbits up to merger.
These are among the longest BNS simulations ever performed, and some of the few 
where an error analysis is available~\cite{Bernuzzi:2011aq,Bernuzzi:2012ci}. 
For each NR data set, we compute the binding energy per reduced mass, 
$E^{\rm NR}_b$, subtracting the GW energy loss from the initial ADM mass,
following~\cite{Damour:2011fu,Bernuzzi:2012ci,Bernuzzi:2013rza}.
Here, differently from previous works, all the multipoles are
included. GW frequency and binding energy are extracted at the moment
of merger. We estimate error bars due to truncation errors and
waveform finite extraction uncertainties from resolution tests for
fewer configurations. More details on these simulations will  
be given elsewhere.  

Recently, Ref.~\cite{Read:2013zra} proposed a phenomenological 
linear relation between the $\log$ of $M\omega_{22\, mrg}^{\rm
  NR}$ and the quantity $\Lambda^{1/5} = (\frac{2}{3}
k_{2})^{1/5}C^{-1}=(\frac{16}{3} \kappa_{2}^{T}(q=1))^{1/5}$
inspecting an independent sample of 
equal-mass, irrotational NR waveforms for six different EOS.  
We believe the effectiveness of that empirical fit is explained by the
$\kappa^{T}_{\ell}$-universality. 

The NR GW frequency $M\omega_{22\, mrg}^{\rm NR}$ and binding 
energy $E^{\rm NR}_{b\, mrg}$ at the moment of merger are plotted as
functions of $\kappa_{2}^{T}$ in Fig.~\ref{fig:MoE}. 
The fit of~\cite{Read:2013zra} complements our numerical data, with
which is perfectly consistent. As indicated by the figure, the NR
points are compatible with the
$\kappa^{T}_{\ell}$-universality. Similarly to the EOB quantities, the 
NR data can be fitted to rational polynomials. We constrain the fit to
the ``black-hole limit''  by factoring out the 
values $E_{b\, mrg}^{\rm NR}(\kappa^T_\ell=0)\approx-0.120$ and
$M\omega_{22\, mrg}^{\rm NR}(\kappa^T_\ell=0)\approx0.360$ as given by
equal-mass binary black hole simulations~\cite{Damour:2011fu}.
The fitting fuction is 
$f(\kappa)=f(0)(1+n_{1}\kappa+n_{2}\kappa^{2})/(1+d_{1}\kappa)$, 
with 
$(n_1,n_2,d_1)=(2.59\cdot10^{-2},-1.28\cdot10^{-5},7.49\cdot10^{-2})$ 
for the frequency and 
$(n_1,n_2,d_1)=(2.62\cdot10^{-2},-6.32\cdot10^{-6},6.18\cdot10^{-2})$
for the binding energy. 
Considering $E_b(\kappa)$ and $M\omega_{22}(\kappa)$ as a parametric
curve, one obtains a relation between the binding energy and the
frequency at the moment of merger that is essentially
linear,
\be
E_{b\, mrg}^{\rm NR} \approx -0.284\, M\omega_{22\, mrg}^{\rm
  NR}-0.0182 \ ,
\ee
with $M\omega_{22\, mrg}^{\rm NR}\in[0.1,0.360]$. Also in this case the black hole
limit is incorporated in the fit. 
Quantitatively, there are differences between the NR merger quantities
$(M\omega_{22\, mrg}^{\rm NR}, E^{\rm NR}_{b\, mrg})$, and the
corresponding EOB ones, $(M\omega_{22\, mrg}^{\rm EOB}, E^{\rm EOB}_{b\, mrg})$, 
see Fig.~\ref{fig:MoE}. 
The relative difference on the relevant interval
$\kappa_2^T\in[50,350]$ is between $20-30\%$ for the frequency and
$10-20\%$ for the binding energy. 
This quantitative disagreement is not surprising: hydrodynamics effects
and nonlinear tidal interactions are not modeled in $A^T(u)$.  At an
effective level, the (uncalibrated) EOB 4PN tidal dynamics basically 
underestimates attractive effects and gives a larger (smaller)
frequency (binding energy) at merger. 
Coincidentally, the adiabatic EOB LSO gives a rather good
numerical approximation, especially for $\kappa_{2}^T\gtrsim200$. 
The key, remarkable point here is
that the adiabatic model already captures the
$\kappa^{T}_{\ell}$-universality, indicating the latter emerges
fundamentally from the conservative dynamics. Furthermore, the simple
LSO analysis gives reasonable estimates of merger relations for any
EOS, mass ratio and (aligned) spins!

\ifusesec
\section{Outlook}
\else
\paragraph*{Outlook.--}
\fi
\label{sec:next}
Modeling GWs from neutron star mergers is a challenging open
problem (see e.g.,~\cite{Wade:2014vqa} for very recent work) 
that can be tackled interfacing accurate nonlinear
simulations with the EOB analytical framework.  
While pursuing this approach we have identified $\kappa^{T}_\ell$
as fundamental ``coupling constants'' of the binary tidal
interactions, together with $\kappa^{T}$-universal relations and their
physical origin. Extension of the present work needs more 
multi-orbit and precise NR simulation including, in particular,
spins~\cite{Bernuzzi:2013rza}. 
Future work will be devoted to explore effective extensions of the
nonadiabatic EOB model, e.g.,~the use of flexibility parameters
or different resummations of $A^T$~\cite{Bini:2012gu}.  
Ultimately, a NR-tuned tidal EOB model is expected to deliver accurate
merger waveforms for BNS GW detection, similar to the black hole
binary case~\cite{Damour:2012ky,Taracchini:2013rva}. 

The $\kappa^{T}$-universality has consequences for GW astronomy.  
For example, using EOB-based merger templates (containing the characteristic
peak) in match filtered searches one might be able to accurately extract the value of
$\kappa_{2}^{T}$ from the template's peak~\cite{Damour:2012yf}. 
A single measure of the frequency at the moment of merger would thus
constrain both the EOS and the binding energy.
The actual possibility to pursue this strategy deserves a study on
its own. 
In this respect, the $\kappa^{T}$-universality characterizing the merger
has similarities with the findings of~\cite{Bauswein:2011tp}
and with the universal relations found for single neutron star
properties~\cite{Yagi:2013bca}. 
Also, we propose to use the value of the merger frequency, as given by our
fits, to mark the end of inspiral templates; this will improve the
simple Schwarzschild LSO criterion, e.g.~\cite{DelPozzo:2013ala}.

Interestingly, due to the 
coincidental ``compensation'' of finite mass effects in the 
tidally dominated regime, the Schwarzschild LSO values 
give very good estimates to the GW frequency and binding energy
at BNS merger for irrotational binaries with $\kappa^T_2\sim200$.

\paragraph*{Acknowledgments.--}
  We thank B.~Br\"ugmann, T.~Damour, C.~Van den Broeck for
  useful comments.   
  This work was supported in part by DFG grant SFB/Transregio~7
  ``Gravitational Wave Astronomy'' and the Graduiertenakademie Jena.
  S.B. and Si.B. thank IHES for hospitality during the development of
  part of this work. Si.B. is supported by the Swiss National Science
  Foundation. M.U. is supported by CAPES under BEX
  10208/12-7, and thanks TPI Jena for hospitality during the
  development of this work. 
  Simulation were performed on the LRZ cluster in
  M\"unich.

\end{document}